\documentclass[fleqn,10pt]{wlscirep}

\usepackage{mathrsfs}
\usepackage{siunitx}

\usepackage{graphicx}
\usepackage{subcaption}
\title{\textit{In situ} phase contrast X-ray brain CT}

\author[1,*]{Linda C. P. Croton}
\author[1,2]{Kaye S. Morgan}
\author[1]{David M. Paganin}
\author[3,4]{Lauren T. Kerr}
\author[3,5]{Megan J. Wallace}
\author[3,5]{Kelly J. Crossley}
\author[3]{Suzanne L. Miller}
\author[6]{Naoto Yagi}
\author[6]{Kentaro Uesugi}
\author[3,5]{Stuart B. Hooper}
\author[1]{Marcus J. Kitchen}
\affil[1]{School of Physics and Astronomy, Monash University, Clayton, Victoria 3800, Australia}
\affil[2]{Chair of Biomedical Physics, Department of Physics, Munich School of Bioengineering, and Institute of Advanced Study, Technische Universit\"at M\"unchen, 85748 Garching, Germany}
\affil[3]{The Ritchie Centre, Hudson Institute of Medical Research, Clayton, Victoria 3800, Australia}
\affil[4]{Cancer Research UK, Angel, London, United Kingdom}
\affil[5]{Department of Obstetrics and Gynaecology, Monash University, Clayton, Victoria 3800, Australia}
\affil[6]{Japan Synchrotron Radiation Research Institute (JASRI/SPring-8), 1-1-1 Kouto, Sayo, Hyogo 679-5198, Japan}

\affil[*]{linda.croton@monash.edu}

\begin{abstract}

Phase contrast X-ray imaging (PCXI) is an emerging imaging modality that has the potential to greatly improve radiography for medical imaging and materials analysis. PCXI makes it possible to visualise soft-tissue structures that are otherwise unresolved with conventional CT by rendering phase gradients in the X-ray wavefield visible. This can improve the contrast resolution of soft tissues structures, like the lungs and brain, by orders of magnitude. Phase retrieval suppresses noise, revealing weakly-attenuating soft tissue structures, however it does not remove the artefacts from the highly attenuating bone of the skull and from imperfections in the imaging system that can obscure those structures. The primary causes of these artefacts are investigated and a simple method to visualise the features they obstruct is proposed, which can easily be implemented for preclinical animal studies. We show that phase contrast X-ray CT (PCXI-CT) can resolve the soft tissues of the brain \textit{in situ} without a need for contrast agents at a dose $\sim$400 times lower than would be required by standard absorption contrast CT. We generalise a well-known phase retrieval algorithm for multiple-material samples specifically for CT, validate its use for brain CT, and demonstrate its high stability in the presence of noise.

\end{abstract}
\begin{document}

\flushbottom
\maketitle

\thispagestyle{empty}

\section*{Introduction}

Several imaging modalities are used for the detection and monitoring of abnormalities within the human brain, most commonly Magnetic Resonance Imaging (MRI) and Computed Tomography (CT). While MRI is the preferred clinical modality for imaging the anatomy of the brain, due to its lack of ionising radiation and a higher contrast resolution\cite{Rankin2008}, CT plays an essential role in diagnostic imaging due to a much faster acquisition speed, which reduces movement artefacts, and a higher spatial resolution. In addition, CT can be used in the presence of metallic implants and tattoos with metallic ink, for which MRI is contraindicated due to the dangers imposed by the high magnetic fields required. CT is also better suited for the evaluation of penetrating brain injuries and other trauma and acute neurological emergencies\cite{Temple2015, Brody2015}, where impromptu access to MRI scanners for time-critical assessment is often unavailable, and the presence of magnetic materials cannot be determined prior to the scan.

X-ray absorption has provided the contrast for computed tomography (CT) since the inception of CT in the 1960s and early 1970s\cite{Cormack1963, Hounsfield1973, Ambrose1973}, enabling the reconstruction of internal `slices' of the human body and of other objects from a series of external views. While absorption CT is very well-suited to imaging soft tissue and bone, its use for delineating more subtle features within the tissues is more limited. Phase contrast X-ray imaging (PCXI) allows us to exploit the diffraction of X-rays, rather than their attenuation, resolving features in soft tissues and other low-density, low-Z materials for which standard attenuation imaging is insufficient. The refractive index decrement, which describes the phase change of X-rays as they pass through matter, is typically three orders of magnitude smaller than the attenuation coefficient for the case of soft tissue in the diagnostic X-ray regime\cite{Cloetens1999, Fitzgerald2000}; this suggests that a significant increase in contrast resolution can potentially be obtained with PCXI. It has recently been shown that phase contrast can enable a radiation dose reduction in CT by factors in the thousands over conventional CT without loss in image quality\cite{Kitchen2017}.

The highest spatial resolution currently achieved for \textit{ex vivo} MRI imaging of a brain is around \SI{50}-\SI{100}{\um} when using extremely high field strengths on the order of \SI{9}-\SI{16}T\cite{Foxley2015} (for reference, clinical CT scanners operate up to \SI{3}T). The best \textit{in vivo} MRI resolution achieved to date is \SI{100}{\um} using \SI{7.00}-\SI{11.7}T scanners\cite{Stucht2015, Wu2016}. To achieve these resolutions, acquisition times must be quite long, typically 1-2 hours. For synchrotron-based phase contrast computed tomography, the highest spatial resolution achieved to date is $\sim$\SI{1}{\um}\cite{Astolfo2016}. While prior work has visualised the brain in mouse fetuses\cite{Hoshino2012}, to date no \textit{in vivo} or \textit{in situ} phase contrast X-ray CT (PCXI-CT) of a brain within a substantially calcified skull has been published, which is likely due to the significant artefacts that typically arise from the skull upon back-projection.

\begin{figure}[h!]
  \centering
    \includegraphics[width=\textwidth]{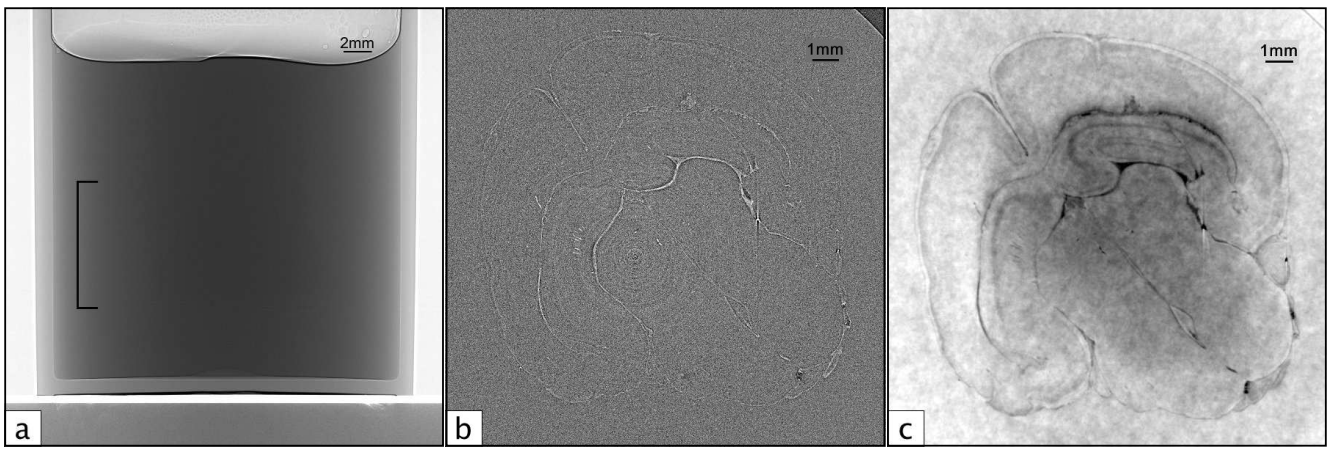}
   \caption{a) A single propagation-based projection image of an excised close-to-term rabbit kitten brain, suspended in agarose. The bracket indicates the position of the brain in the tube. b) Phase contrast tomographic slice of the rabbit kitten brain in (a). Note that the visible structures can only be seen due to the phase shifts imparted by propagation and would not be visible at all in absorption contrast CT. c) Phase-retrieved tomogram of the brain from (a) and (b). CTs were acquired at 24 keV with an object-to-detector distance of 5 m. See methods section for experimental details.}
  \label{fig:bunnybrain}
\end{figure}

Recent studies have demonstrated that PCXI-CT is an effective tool for small animal studies of the brain, providing high resolution images of tissue structures and clear delineation between grey and white matter\cite{Pfeiffer2007,Schulz2010,Beltran2011,Pinzer2012,Huang2015}. Beltran et al.\cite{Beltran2010,Beltran2011} showed a 200-fold increase in signal-to-noise ratio using PCXI-CT over absorption contrast, indicating that \textit{in situ} PCXI-CT can lead to very large improvements over conventional absorption contrast CT. A similar example from our data collected at the SPring-8 synchrotron in Japan is shown in Fig. \ref{fig:bunnybrain}, showing the signal-to-noise ratio (SNR) and contrast-to-noise ratio (CNR) improvement that can be achieved with phase contrast and phase retrieval.

These previous studies have all been limited to \textit{ex vivo} imaging on brains that have been excised from their skulls. While these results show the clear potential for brain PCXI-CT in preclinical studies, \textit{in vivo} imaging is much more difficult due to the strong phase gradients between tissue and bone as well as the strong absorption by bone, causing artefacts from the skull which obscure structures that would otherwise be well-resolved. Overcoming these artefacts is important for future \textit{in vivo} preclinical research using imaging and, ultimately, for adaptation to the clinic. Herein we demonstrate the first visualisation of the brain \textit{in situ} in a small animal model, performed using propagation-based PCXI-CT.

\section*{Background}

\subsection*{Propagation-based phase contrast imaging}

Propagation-based imaging (PBI) is the simplest phase-contrast method, wherein the phase shifts resulting from refraction within the object are converted to intensity variations via propagation between the object and the detector, with no optical components required along the path. As the wavefront propagates, small differences in phase accumulate between contrasting materials and/or changing thicknesses so that Fresnel fringes become clearly visible at the detector. The experimental setup for PBI differs from conventional X-ray absorption imaging only in the distance between the imaged object and the detector and in the requirement of a source with sufficiently high spatial coherence\cite{Cloetens1996, Wilkins1996}.

\begin{figure}[h!]
  \centering
    \includegraphics[width=\textwidth]{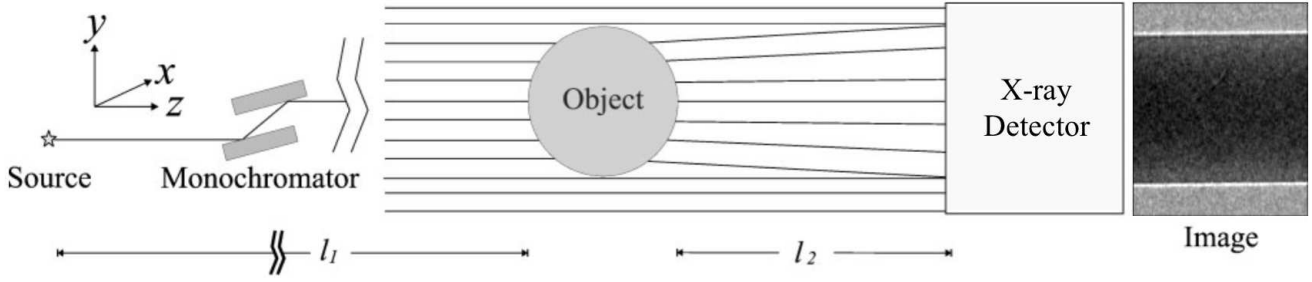}
    \caption{Propagation-based X-ray phase contrast imaging (adapted from Kitchen et al., 2005\cite{Kitchen2005}).}
  \label{fig:pbi}
\end{figure}

As incident X-rays pass through a sample, the intensity measured downstream at the detector contains a combination of attenuation and phase information. In Fig. \ref{fig:pbi}, a simple, single-material sample is shown with the resultant image at the detector exhibiting contrast that is mostly due to attenuation in the regions within and surrounding the sample; however the phase effects are clear at the boundaries between materials of differing projected refractive index (in this case, the sample material and the surrounding air). This results in the fringe pattern discussed above, seen on the right side of the figure. This fringe pattern can be used to extract quantitative information about the sample, as described in the next section.

\subsection*{Simplified phase retrieval}

The behaviour of an X-ray wavefield, as it passes through an object, is governed by the complex refractive index of the sample, which, for a monochromatic incident wavefield, ignoring polarization, is given by,
\begin{equation}
  n(\mathbf{r}) = 1 - \delta(\mathbf{r}) + \textit{i}\beta(\mathbf{r}).
\end{equation}
At each point $\mathbf{r}$ in the sample, the real part $\delta$ is known as the refractive index decrement, describing the phase component, and $\beta = \lambda\mu/4\pi$ is the attenuation index, describing the absorption component, where $\lambda$ is the wavelength and $\mu$ is the attenuation coefficient of the material. 

Under the projection approximation, which assumes that the transverse scattering contribution to the deflection of X-rays through a sample is negligible, the intensity immediately downstream of the sample (i.e. the detector side of the sample) is given by Beer's Law,
\begin{equation}
  I(\mathbf{r}_\bot) = I_0 e^{-\int \mu(r_\bot,z) dz},
  \label{eq:beers}
\end{equation}
where $I_0$ is the intensity of the monochromatic plane waves assumed to be incident on the sample, $z$ is the direction of propagation, and $\mathbf{r}_\bot$ represents the coordinates perpendicular to propagation. Similarly, the phase is given by,
\begin{equation}
  \phi(\mathbf{r}_\bot) = -k \int \delta(\mathbf{r}_\bot,z) dz.
\end{equation}
For a single-material object, these become,
\begin{equation}\label{eq:intensityphase}
I(\mathbf{r}_\bot) = I_0 e^{-\mu T(\mathbf{r}_\bot)}\ \ \mathrm{and}\ \ \phi(\mathbf{r}_\bot) = -k\delta T(\mathbf{r}_\bot),
\end{equation}
where $T(\mathbf{r}_\bot)$ is the projected thickness of the sample. Since the intensity of a propagating wave is $I(\mathbf{r}) = | \psi(\mathbf{r})|^{2}$, the propagating wavefield $\psi$ can be represented as:
\begin{equation}
  \psi (\mathbf{r}) = \sqrt{I(\mathbf{r})}e^{i\phi (\mathbf{r})}.
\end{equation} 
In the near-field regime (Fresnel number $N_F \gg 1$), the transport-of-intensity equation\cite{Teague1983} holds:
\begin{equation} \label{eq:tie}
  \nabla_\bot \cdot [I(\mathbf{r})\nabla_\bot \phi(\mathbf{r})] = -k\frac{\partial{I(\mathbf{r})}}{\partial{z}}.
\end{equation}
Here, $\nabla_\bot$ denotes the gradient operator in the $x-y$ plane perpendicular to the optic axis $z$.

From this equation and from the intensity and phase given above under the projection approximation, Paganin et al.\cite{Paganin2002} derived the following expression to recover the projected thickness for a single-material sample,
\begin{equation}\label{eq:Paganin2002}
  T(\mathbf{r}_\bot) = -\frac{1}{\mu} \log_e \left( \mathscr{F}^{-1} \left\{ \mu \frac{ \mathscr{F} \left\{ I(\mathbf{r}_\bot,z=R_2)/I_0 \right\}}{R_2 \delta |\mathbf{k_{\bot}}|^2 + \mu} \right\} \right),
\end{equation}
where $I(\mathbf{r}_\bot ,z=R_2)$ is the intensity at the detector plane and $R_2$ is the sample-to-detector distance. Equation (\ref{eq:Paganin2002}) was expanded upon by Beltran et al.\cite{Beltran2010} for the case of a two-material sample, giving:
\begin{equation} \label{eq:beltranthickness}
T_2 (\mathbf{r}_\bot) = -\frac{1}{\mu_2 - \mu_1} \times \ \log_e \left( \mathscr{F}^{-1} \left\{ \frac{1}{[R_2 (\delta_2 - \delta_1)/(\mu_2 - \mu_1)]\mathbf{k}_{\bot}^2 + 1} \mathscr{F} \left[ \frac{\mathrm{I}(\mathbf{r}_\bot,z=R_2)}{I_0 e^{-\mu_1 A(\mathbf{r}_\bot)}} \right] \right\} \right),
\end{equation}
where the subscripts $1$ and $2$ refer to the two different materials, the first embedded within the second, and $A(\mathbf{r}_\bot)$ is the combined projected thickness, $A(\mathbf{r}_\bot) = T_1(\mathbf{r}_\bot) + T_2(\mathbf{r}_\bot)$. Determining $A(\mathbf{r}_\bot)$ is not easy, or may only be possible with limited accuracy, depending on the sample. Instead, we diverge slightly from their method, as outlined below.

Recovery of the exit surface intensity of an object from projection images of one material embedded within another can be achieved by modifying the work of Beltran et al.\cite{Beltran2010} Their equation (8) is 
\begin{equation} \label{eq:beltran}
\frac{\textrm{I} (\mathbf{r}_{\bot}, z = R_2)}{I_0 \exp{\left[ -\mu_1 A(\mathbf{r}_{\bot})\right]}} = \left[-\frac{R_2 (\delta_2 - \delta_1)\nabla^{2}_{\bot}}{(\mu_2 - \mu_1)} + 1 \right] \exp{\left[ -(\mu_2 - \mu_1)T_2(\mathbf{r}_\bot) \right]}.
\end{equation}
Here
\begin{equation} \label{eq:totalthickness}
A(\mathbf{r}_\bot) = T_1(\mathbf{r}_\bot) + T_2(\mathbf{r}_\bot).
\end{equation}
Since $A(\mathbf{r}_\bot)$ is assumed to be a slowly varying function of $\mathbf{r}_\bot$, the denominator of the left-hand side of equation (\ref{eq:beltran}) can be approximated as a constant when a spatial derivative such as $\nabla^{2}_{\bot}$ is applied to it. When both sides of equation (\ref{eq:beltran}) are multiplied by this term, it can therefore be grouped with the exponential term on the right-hand side to give
\begin{equation}
I_0 \exp{\left[ -\mu_1 A(\mathbf{r}_\bot) \right]} \exp{\left[ -(\mu_2 - \mu_1) T_2(\mathbf{r}_\bot) \right]}  =   I_0 \exp{\left[ -\mu_1 T_1(\mathbf{r}_\bot) - \mu_2 T_2(\mathbf{r}_\bot) \right]}   =   I_0 \textrm{I}(\mathbf{r}_{\bot}, z = 0).
\end{equation}
Therefore, we can rewrite equation (\ref{eq:beltran}) as
\begin{equation}
\frac{\textrm{I} (\mathbf{r}_{\bot}, z = R_2)}{I_0} = \left[- \frac{R_2 (\delta_2 - \delta_1) \nabla^{2}_{\bot}}{(\mu_2 - \mu_1)} + 1 \right] \textrm{I} (\mathbf{r}_{\bot}, z = 0).
\end{equation}

An alternative approach is to consider that, for a two-material sample under the projection approximation, equations (\ref{eq:intensityphase}) become,
\begin{equation}
I(\textbf{r}_\bot) = I_0 \exp{\left[ -\mu_1 T_1(\textbf{r}_\bot) - \mu_2 T_2(\textbf{r}_\bot)\right]}\ \ \mathrm{and}\ \ \phi(\mathbf{r}_\bot) = -k\delta T_1 (\mathbf{r}_\bot) -k\delta T_2 (\mathbf{r}_\bot).
\end{equation}
Substituting for $T_1 (\mathbf{r}_\bot)$ using equation (\ref{eq:totalthickness}) gives,
\begin{equation}
I(\mathbf{r}_\bot) = I_0 \exp{\left[ -\mu_1 A(\mathbf{r}_\bot) \right]} \exp{\left[ -(\mu_2 - \mu_1) T_2 (\mathbf{r}_\bot) \right]}\ \ \mathrm{and}\ \ \phi(\mathbf{r}_\bot) = -k\delta_1 A(\mathbf{r}_\bot) - (\delta_2 - \delta_1) T_2 (\mathbf{r}_\bot).
\end{equation}
Assuming, again, that $A(\mathbf{r}_\bot)$ can be approximated as a constant, this results in a multiplicative change in the measured intensity and an additive shift in the phase, which disappears on differentiation in equation (\ref{eq:tie}); hence, this becomes like a single-material object where $\delta = \delta_2 - \delta_1$ and $\mu = \mu_2 - \mu_1$ in equation (\ref{eq:Paganin2002}).

Following Paganin et al.\cite{Paganin2002}, Beltran et al.\cite{Beltran2010}, and Gureyev et al.\cite{Gureyev2002}, the ``absorption" contrast image can be recovered using the Fourier derivative theorem:
\begin{equation}\label{eq:twomaterials}
\textrm{I}(\textbf{r}_{\bot}, z = 0) = \mathscr{F}^{-1} \left\{ \frac{1}{\left[ R_2 (\delta_2 - \delta_1)/(\mu_2 - \mu_1) \right] \mathbf{k}^{2}_{\bot} + 1} \mathscr{F} \left[ \frac{\mathrm{I} (\mathbf{r}_{\bot}, z = R_2)}{I_0} \right] \right\}.
\end{equation}
A tomographic tilt series of such projections can then be used to quantitatively reconstruct the objects embedded within the encasing material. A similar approach can be used to solve for the encasing material itself by following the algorithm of Paganin et al. We see that our revised equation does not require knowledge of the total projected thickness $A(\mathbf{r}_\bot)$, unlike the algorithm of Beltran et al. (equation (\ref{eq:beltranthickness})). We also note that, for tomography, we need only recover the absorption contrast image before using a tomographic reconstruction algorithm. In the slightly different context of phase retrieval of several sharp boundaries in tomography, Haggmark et al.\cite{Haggmark2017} recently came to the same equation, also utilising the assumption of a slowly-varying thickness in addition to assuming a linear relationship between $\delta$ and $\mu$ for different materials at a given energy.

\subsection*{Single image phase retrieval of a brain \textit{in situ}}

\begin{figure}[h!]
  \centering
    \includegraphics[width=\textwidth]{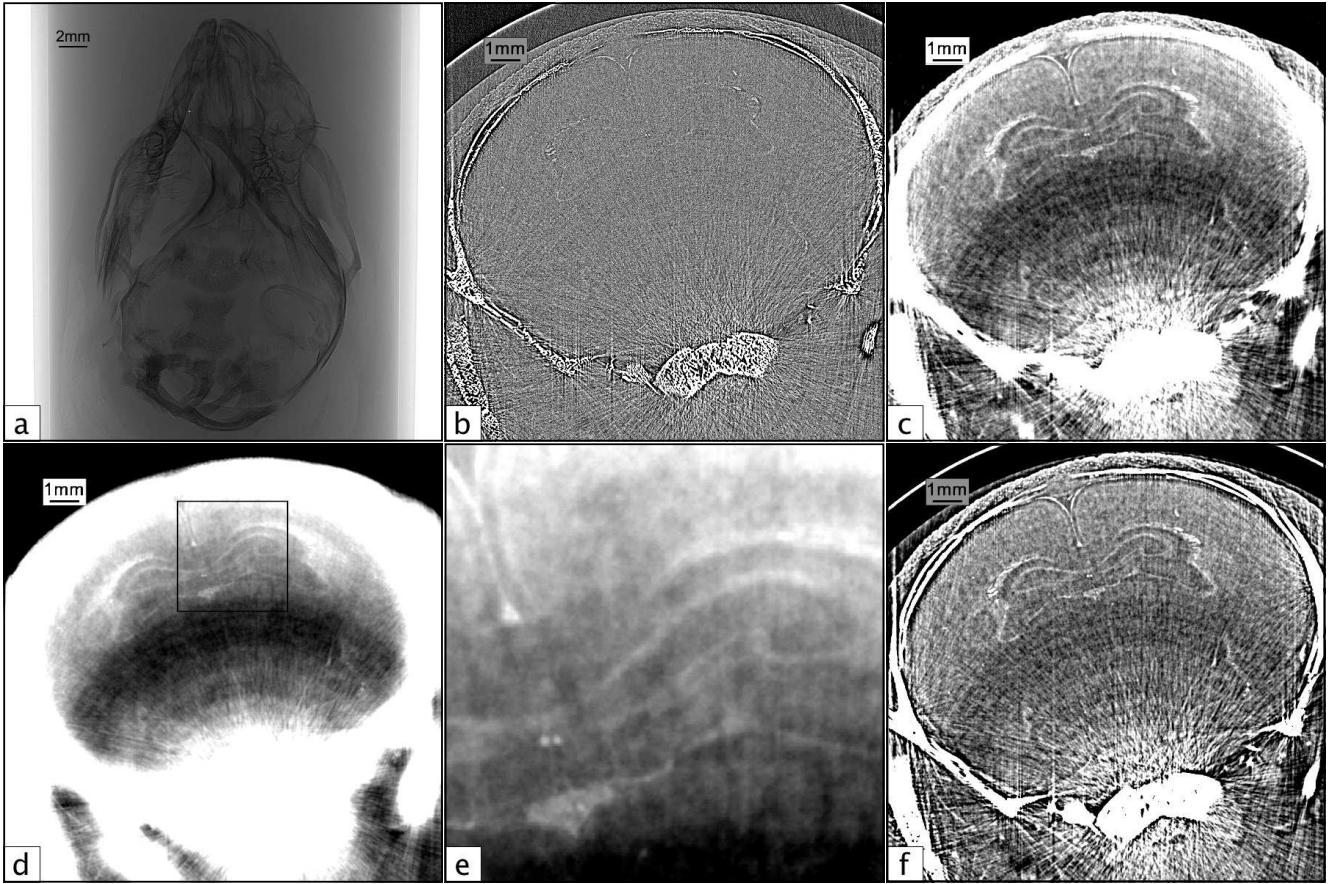}
  \caption{Panels (a) - (c) are the \textit{in situ} analogue to the panels in Fig. \ref{fig:bunnybrain}(a) - \ref{fig:bunnybrain}(c) for a full head of a dead close-to-term rabbit kitten, though the region of the brain is not the same as Fig. \ref{fig:bunnybrain}. CTs were acquired at 24 keV with a propagation distance of 5 m. a) A single propagation-based projection image of a whole rabbit kitten head suspended in agarose. b) Phase contrast tomogram of the rabbit kitten head in (a). c) Phase retrieved tomogram of the head from (a) and (b) created using the single-material algorithm with respect to a bone/air interface. d) Phase retrieved tomogram using the single-material algorithm with respect to a tissue/air interface. e) A close-up of the square region outlined in (d). f) Phase retrieved tomogram using the two-material algorithm (equation (\ref{eq:twomaterials})) with respect to the bone/tissue interface.}
  \label{fig:bunnyhead}
\end{figure}

For materials of similar composition (and hence refractive index), such as the grey and white matter of the brain, the single-material phase retrieval algorithm applied with respect to the tissue/air interface works quite well to resolve the boundaries between these materials. Structures that are unresolved or poorly resolved with attenuation contrast alone become visible upon phase retrieval, since the phase retrieval filter suppresses noise, thereby increasing the SNR and CNR. Ideally, phase retrieval performed with respect to the grey/white matter boundary provides the best contrast resolution of brain structures, however when imaging the brain \textit{in situ}, the inaccurate phase retrieval at the bone/tissue interface causes excessive blurring that overwhelms the features within the soft tissue of the brain. Fig. \ref{fig:bunnyhead} shows the \textit{in situ} analogue to Fig. \ref{fig:bunnybrain}. In panel \ref{fig:bunnyhead}c, phase retrieval has been performed with respect to bone/air interface, resulting in an over-blurring at both the grey/white matter and bone/tissue boundaries. Nevertheless, the brain structure is more clearly resolved in Fig. \ref{fig:bunnyhead}c than in Fig. \ref{fig:bunnyhead}b, despite the incorrect phase retrieval at the bone/tissue interface and the associated artefacts caused by the the highly attenuating bone. In panel \ref{fig:bunnyhead}d, phase retrieval has been performed with respect to the tissue/air interface, again resulting in a highly-blurred reconstruction. In this case, brain features are very well resolved (panel \ref{fig:bunnyhead}e) but are dominated by the bone artefacts.

When phase retrieval is performed with respect to the bone/tissue interface using equation (\ref{eq:twomaterials}) ($\Delta\delta/\Delta\mu = (\delta_2 - \delta_1) / (\mu_2 - \mu_1) = 5.66 \times 10^{-10}$), both interfaces are better resolved with a more consistent resolution across the image than with the single-material algorithm ($\delta/\mu = 1.54 \times 10^{-9}$ and $\delta/\mu = 8.60 \times 10^{-9}$ for bone and brain tissue, respectively). These results are shown in panel \ref{fig:bunnyhead}f, where it can be seen that the artefacts due to the highly absorbing bone are largely abated. The remaining artefacts fall into two types - ring artefacts and streak artefacts - and are ones which should ideally be corrected for prior to the phase-retrieval step. Streak artefacts across high-contrast edges are caused by multiple factors, and determining the dominant contributing factors is important for improving the SNR and CNR for \textit{in situ} brain imaging.

\subsection*{Streak artefacts in CT}
\label{sec:artefacts}

Numerous phenomena contribute to the formation of streak artefacts in CT reconstruction, particularly along high contrast edges. The three main causes are 1) insufficient data within the dynamic range of the detector, 2) noise, 3) physical effects that reduce contrast and resolution. These all create strong artefacts when there are large attenuation gradients within the sample, particularly when trying to detect subtle variation in soft-tissue contrast (e.g. grey/white matter). For a brain \textit{in situ}, artefacts from the bone that are not usually problematic become important, since they can overwhelm the underlying, relatively low-contrast brain structures. Even a sub-pixel offset in the centre of rotation correction can create obvious tuning-fork artefacts\cite{Shepp1979} from the bone that are not apparent when viewing an intensity palette that is scaled to include higher-density structures and hence a larger contrast range. The following phenomena are the primary contributors to streak artefacts in CT:

\begin{itemize}

\item \textit{Photon starvation} - This phenomenon occurs when an insufficient number of photons reach the detector through a highly-attenuating region of a sample, such as a metal implant or bone. When this occurs, the signal at that part of the detector is dominated by noise, leading to streak artefacts on reconstruction. There are a number of methods employed to correct for these artefacts, many of which involve thresholding and interpolation of the sinogram. These methods work quite well for compact regions where the region can easily be removed by thresholding in the sinogram without interfering significantly with neighbouring regions (for a discussion of these techniques, see Mouton et al.\cite{Mouton2013}).

\item \textit{Beam Hardening} - The relationship between the logarithm of the X-ray intensity transmitted through a sample and the sample thickness is linear at any given energy (Beer's Law) under the projection approximation; however, this is only strictly true for a monochromatic source. Lower-energy photons are absorbed more readily than higher-energy photons, leading to a deviation from this linearity for a polychromatic source (i.e. beam hardening). This change in the attenuation profile from the monochromatic case results in cupping artefacts through individual materials and streak artefacts along edges between different materials in the tomographic reconstruction.

\item \textit{Energy Harmonics} - For synchrotron sources, monochromators are used to limit transmission of the source to a narrow band of energies.  Energies outside the desired value can cause a form of beam hardening if the spectral bandwidth is too broad or if higher-order harmonics of the monochromator crystal are allowed to pass through. In the former case, a standard beam-hardening correction can be applied, while the latter requires a correction that accounts for the specific energies associated with the higher order harmonics.

\item \textit{Compton scattering}\label{sec:cs} - Compton scattering occurs as X-rays pass through the sample and can contribute to streak artefacts by increasing the background intensity, hence decreasing the contrast across edges. The scattering cross-section is lobed in the forward direction but amounts to a relatively uniform contribution for the energies and fields of view typically used for small animal imaging. This effect diminishes on propagation and is expected to have a small effect for PBI imaging at object-to-detector propagation distances over 2 m.

\item \textit{Poisson Noise} - Noise in an imaging system has a similar effect to Compton scattering, with the exception that noise exhibits substantial spatial variations while Compton scattering is typically spatially smooth. Like Compton scattering, it will vary the background signal, resulting in a change in the attenuation gradient across high-contrast boundaries. This leads to an over- or under-estimation of the attenuation in the regions tangential to these edges in the CT acquisition plane.

\item \textit{Point Spread Function} - Imperfect detector response plays a significant role in streak artefacts, since it results in a decrease in image resolution, leading to blurring across edges. Deconvolution of the point spread function sharpens the edges, but since it also amplifies noise, there is a tradeoff in doing so. This is covered in more detail in the results section under `simulation and experiment'.

\item \textit{Edge Effects} - The discrete nature of detector pixels means that there is a spatial averaging of the signal at each pixel when a sharp edge lies within the boundaries of a pixel\cite{Joseph1981}. The relevance of this effect increases with increasing pixel size, so it can be minimised with the use of high-resolution detectors. Edge effects are also tied in with the detector PSF and can therefore be accounted for, to some degree, in the process of deconvolution.

\item \textit{Centre of rotation offset} - When acquiring a parallel-beam CT, it is nearly impossible to position the sample such that the centre of rotation corresponds exactly to the centre or edge of a pixel, resulting in an offset that must be accounted for when reconstructing. This is relatively easy to do, and there exist a number of different methods to determine the offset (for discussion of some of these methods, see Vo et al.\cite{Vo2014}). In most cases, it is sufficient to determine this offset to the nearest pixel. When it is not sufficient, the effect can be minimised to a local blurring, rather than streaks, by acquiring projections across 360$^{\circ}$ rather than 180$^{\circ}$. Alternatively, we find that by mirroring the projections of a 180$^{\circ}$ CT and reconstructing the volume as if it were acquired over 360$^{\circ}$, the same effect can be achieved.

\end{itemize}

Compton scattering can be ruled out as a significant contributor to the streak artefacts in our experiments, since the scattered photon density is both proportional to the sample size and inversely proportional to the object-to-detector distance\cite{Sorenson1985}. We expect the scatter-to-primary ratio to be less than a few percent due to the very small beam size (3 cm $\times$ 3 cm) and large propagation distance (5 m) at relatively low energy (24 keV). Several of the other effects described above are discussed further in the results section.

\subsection*{Ring artefacts}
\label{sec:rings}

Synchrotron CT is also subject to ring artefacts, which are due to temporal variations in the intensity of the beam and possibly also non-linearities in the detector response, both of which prevent proper correction using a `flat-field' image. Minimising ring artefacts is particularly challenging for \textit{in situ} brain imaging, since many ring removal methods exploit their circular symmetry (see M\"unch et al.\cite{Muench2009} and Prell et al.\cite{Prell2009} for examples); since the skull is also largely circularly symmetric, this makes it difficult to decouple signal from artefact, resulting in an over-correction of some skull regions, where sections of bone are interpreted as artefacts, and an associated under-correction of regions that are diametrically opposite. While the development of effective ring artefact correction algorithms is essential for preclinical studies of the brain, it is omitted here as beyond the scope of this work; however it should be noted that a better option may be to prevent these artefacts from forming in the first place by shifting the sample in regular manner with respect to the detector (e.g. by translating the sample or detector vertically) during acquisition of the CT. This would prevent the variations in pixel sensitivity that are responsible for the ring artefacts from being consistently present in a single reconstruction plane. Another possibility is to address the problem where it initiates by more accurately characterising the detector components, as well as any time fluctuations in the X-ray source, in order to properly account for them.

\section*{Results}
\label{sec:results}

\subsection*{Streak artefacts from limited spatial resolution: simulation and experiment}
\label{sec:phantomandsim}

\begin{figure}[h!]
  \centering
    \includegraphics[width=\textwidth]{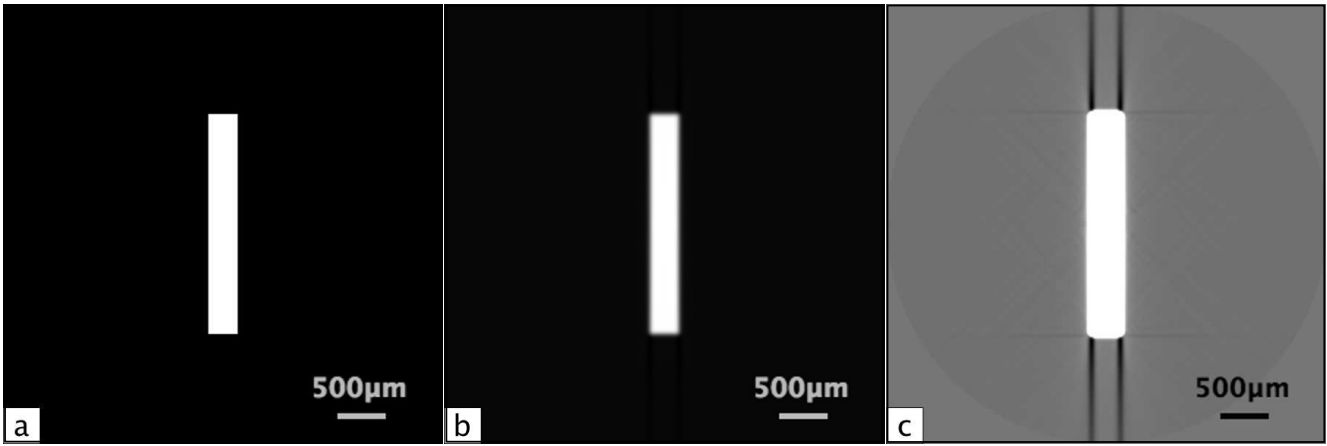}
  \caption{a) A simulated absorption contrast CT of an aluminium rod surrounded by air. b) The same rod from (a) reconstructed from projections that were blurred with a 2D Gaussian. c) The same blurred rod from (c) with the palette scaled to focus on features in the background. The simulation was designed to have similar projected thicknesses and length to the aluminium phantom.}
  \label{fig:simulation}
\end{figure}

To explore the effects that the detector PSF has on CT imaging, an absorption contrast CT of an aluminium rod in air was simulated, assuming an X-ray source energy of 26 keV and using the corresponding attenuation coefficient for aluminium, $\mu = 445.23m^{-1}$. The projection images were then convolved with a 2D Gaussian before back projection. The simulation results are shown in Fig. \ref{fig:simulation}, where panel \ref{fig:simulation}a is the initial ideal object and panel \ref{fig:simulation}b is the resulting reconstruction after PSF blurring. The latter is still a reasonable representation of the ideal object when the intensity palette is scaled to display the full intensity range; however, when the scaling is adjusted to highlight the subtle variations in the background region (\ref{fig:simulation}c), dark streak artefacts are clearly evident emanating from the high-contrast edges, with larger effects along longer edges.

\begin{figure}[h!]
  \centering
    \includegraphics[width=\textwidth]{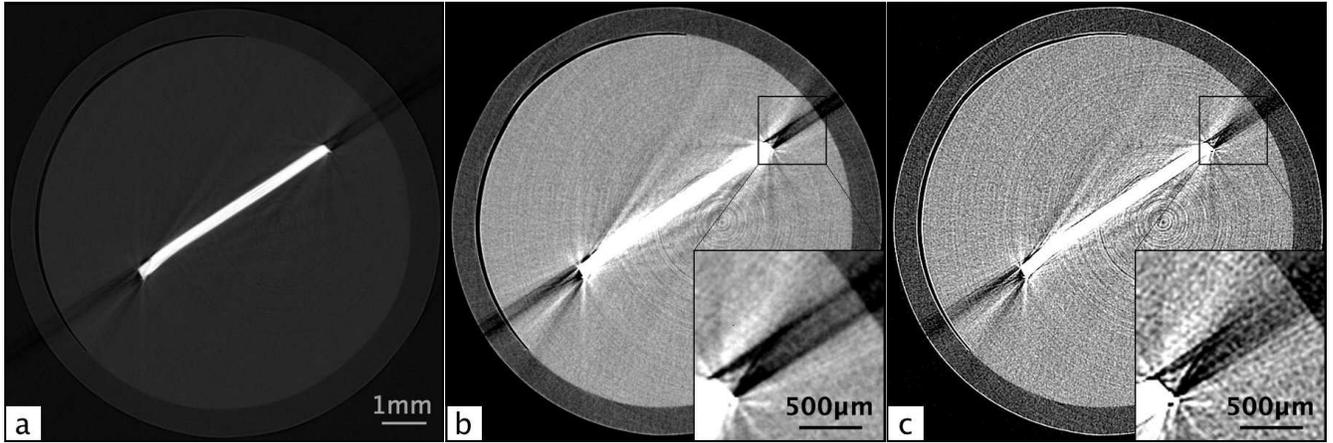}
  \caption{a) A reconstructed slice through a phantom consisting of a 0.3 mm thick aluminium strip suspended in agarose, scaled to the full greyscale palette. b) The same phantom as in (a), with the palette scaled to the background region containing artefacts. c) The same phantom after Wiener deconvolution of the measured point spread function (see the simulation and experiment section for PSF details). The phantom CTs were acquired at 26 keV at a propagation distance of 12 cm.}
  \label{fig:trapezoiddec}
\end{figure}

Figure \ref{fig:trapezoiddec} shows a reconstruction of the aluminium phantom, measuring approximately 5 mm in length in the CT acquisition plane. An energy of 26 keV was used for imaging to ensure sufficient transmission through the relatively dense aluminium. Panel \ref{fig:trapezoiddec}a shows the image scaled to include the full greyscale palette in the same way as the simulation in panel \ref{fig:simulation}b, clearly delineating the aluminium strip with only minimal artefacts. When scaled to the background palette (panel \ref{fig:trapezoiddec}b), the extent of the artefacts is more apparent. In the inset image, dark streak artefacts can be seen that resemble those in the PSF simulation of panel \ref{fig:simulation}c. The detector line spread function was measured to sub-pixel accuracy in two orthogonal directions using a \SI{250}{\um} thick tungsten edge and a Pearson type VII function\cite{Pearson1916}:
\begin{equation}
y = c \left[1 + \frac{(x - x_0)^2}{m a^2} \right]^{-m}.
\end{equation}
The PSF was then approximated by a 2D elliptical Pearson VII function:
\begin{equation}
y = c \Bigg( 1 + \frac{1}{m} \left[ \frac{(x -x_0)^2}{a_h^2} + \frac{(y - y_0)^2}{a_v^2}\right] \Bigg)^{-m},
\end{equation}
where $(x_0, y_0)$ is the centre of the PSF image, $m$ is the larger of the two parameters $m$ from the orthogonal line spread function fits, and $a_h$ and $a_v$ are the amplitudes of the horizontal and vertical fits, respectively. The Pearson type VII function was chosen to represent the PSF, because it ensures that the denominator in the deconvolution is always greater than zero. The exact parameters used for this fit are $c = 0.0701$, $m = 3.222$, $a_h = 1.443$, and $a_v = 1.477$. The PSF was found to be much broader than expected, with visible effects across sharp edges spanning $\sim$80 pixels. The full-width at half-maximum (FWHM) is $\sim$ 2.6 pixels.
When the PSF is deconvolved using Wiener deconvolution, the effect of the streak artefacts is improved, but there is clearly still a large component remaining (panel \ref{fig:trapezoiddec}c), indicating that other effects are involved.

These results show substantial artefacts that dominate the background of the image in the regions immediately surrounding the metal phantom. In many ways, the skull is more forgiving in that its roughly circular symmetry results in an averaging out of many of the streak artefacts in addition to restricting the bulk of them to the region outside its boundary; however, the skull is also the biggest obstacle in correcting those artefacts as it completely surrounds the brain. This fact rules out the use of many of the most common metal artefact reduction techniques available, which often involve some form of thresholding and in-filling of the sinogram, that might otherwise be applied to reduce bone artefacts. Even with as precise as possible thresholding and adaptation of a normalisation method similar to that used by Meyer et al.\cite{Meyer2010}, we find that the skull is too broadly distributed across the sinogram to differentiate it from soft tissue for in-filling. Iterative reconstruction methods, however, may prove to be useful for artefact reduction\cite{Mouton2013}.

\subsection*{Harmonic contamination measurement}

\begin{figure}
  \centering
    \includegraphics[width=0.6\textwidth]{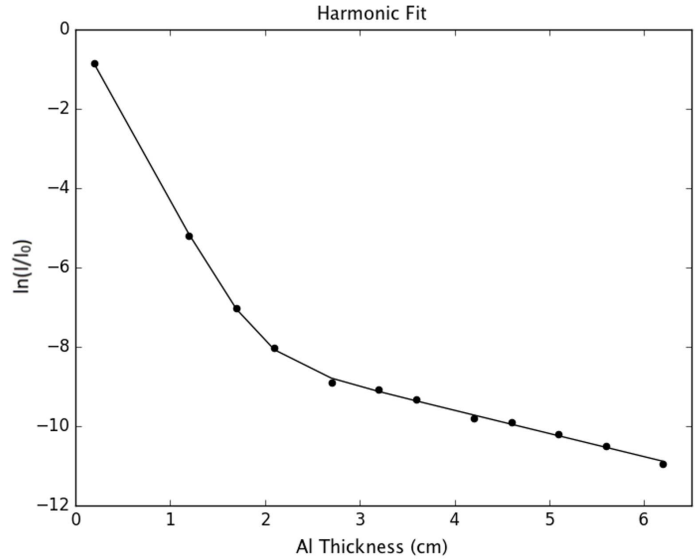}
  \caption{The negative natural logarithm of the normalised intensity as a function of aluminium thickness for a nominally 26 keV
spectrum of synchrotron X-rays filtered by the (111) reflection from a Si monochromator. For a purely monochromatic source, this is expected to be a straight line as per Beer's Law. This deviation from linearity is due to contribution from the third harmonic of the monochromator crystal, which becomes dominant at larger thicknesses. The slope of the curve below $\sim$20 mm corresponds to the attenuation coefficient of Al at the fundamental energy, and at larger thicknesses it corresponds to that of the third harmonic.}
  \label{fig:dlfit}
\end{figure}

Finally, a harmonic contamination test was performed to determine the significance of the monochromator harmonics. Images were acquired with aluminium sheets of increasing thickness placed in front of the detector until the measured signal fell to that of the dark current. The negative natural logarithm of the mean normalised intensity, $-\log_e(I/I_0)$, was plotted as a function of aluminium thickness and fit, using least squares minimisation, as:
\begin{equation}
-\log_e (I/I_0) = -\log_e \left[(1 - a) \exp(-\mu_F T) + a \exp(-\mu_H T)\right],
\end{equation}
where $\mu_F$ and $\mu_H$ are the attenuation coefficients corresponding to the fundamental and third harmonics of the Si (1 1 1) monochromator crystal, $T$ is the thickness of the aluminium, and $a$ is the fractional contribution of the fundamental harmonic to the mean intensity at the detector. The fit can be seen in Fig. \ref{fig:dlfit}, with a value of $a < 1\%$. The contribution of the third harmonic was found, as expected, to be an insignificant contribution to the artefacts created on back projection.

The phenomena outlined in the previous few sections are the most common causes of streak artefacts in CT. These simulations and experiments have shown that the artefacts cannot be easily explained by one individual source. The phenomenon that clearly does contribute - the PSF - is not sufficient to account for the full extent of the artefacts. We conclude that there are likely a number of different phenomena contributing that, while occurring individually might only have a small affect, together amount to a much larger cumulative effect. Isolation and correction of these effects remains the subject of future work.

\subsection*{Rabbit kitten brain CT}

\begin{figure}[h!]
  \centering
    \includegraphics[width=\textwidth]{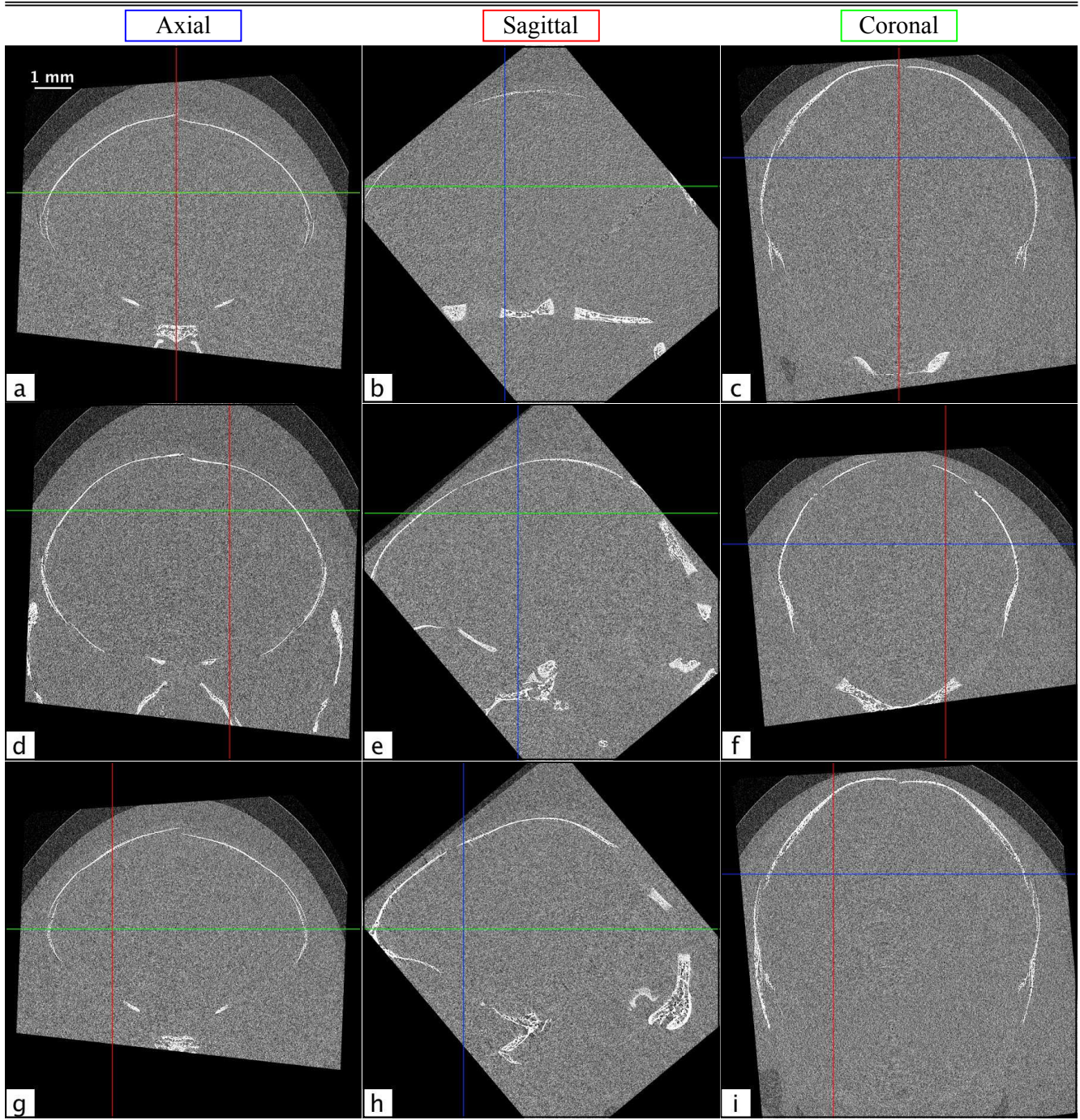}    
  \caption{Several axial (a, d, and g), sagittal (b, e, and h), and coronal (c, f, and i) absorption contrast tomograms from \textit{in situ} brains from dead rabbit kittens. The crosshairs on each image denote the locations of the slices from the other orientations in each row, with axial, sagittal, and coronal slices marked in blue, red, and green, respectively.}
  \label{fig:fadpanel}
\end{figure}

\begin{figure}[h!]
  \centering
    \includegraphics[width=\textwidth]{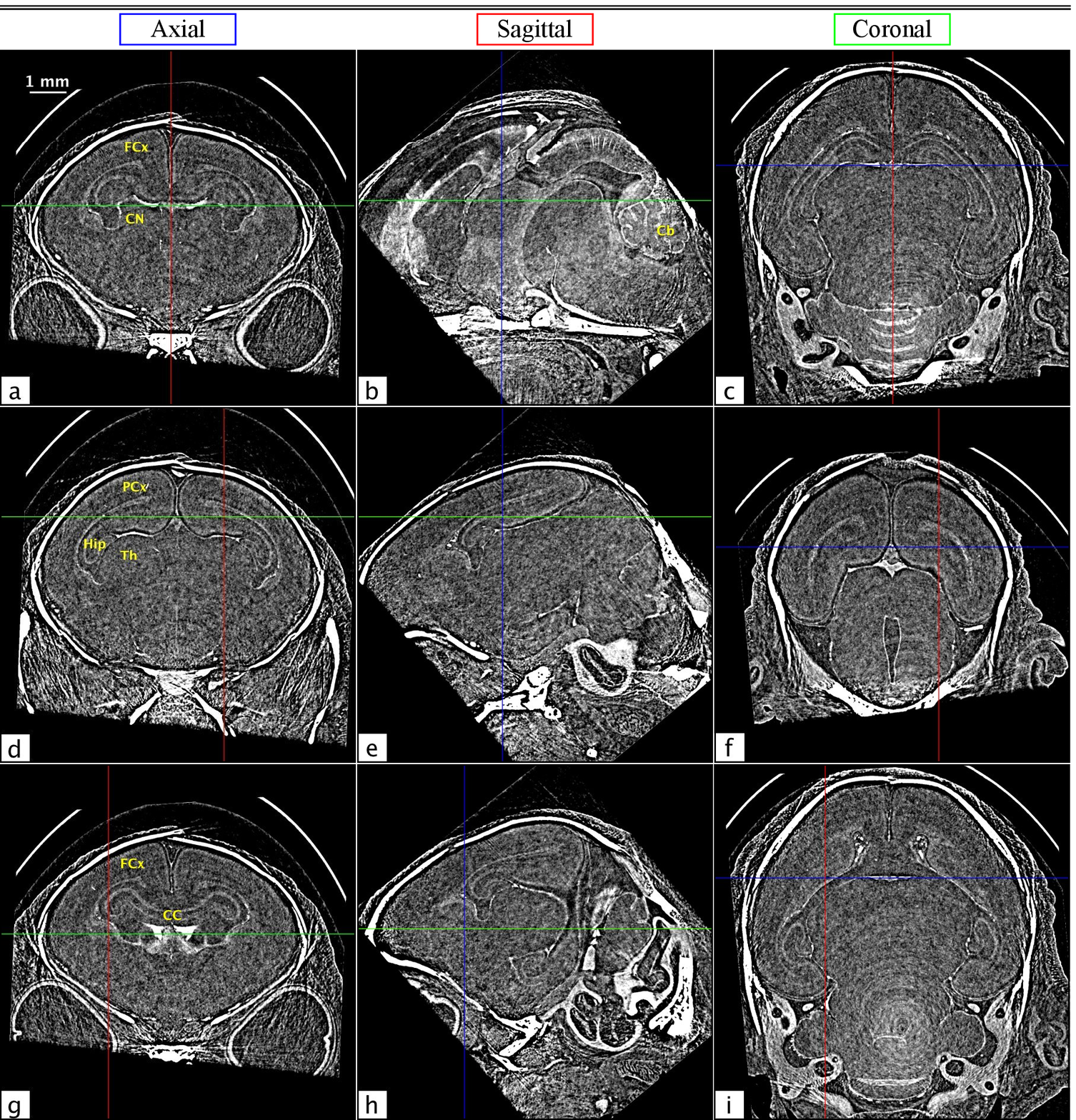}
  \caption{The same slices from Fig. \ref{fig:fadpanel}, now with phase contrast and phase retrieval. As with Fig. \ref{fig:fadpanel}, axial, sagittal, and coronal views are marked in blue, red, and green, respectively. Circularly symmetric ring artefacts can be seen as a white blurring toward the bottom of panels \ref{fig:siprpanel}(c) and \ref{fig:siprpanel}(i). These also manifest as a diffuse white band in panel \ref{fig:siprpanel}(b), running from the lower centre of the image toward the cerebellum (Cb). Note that these artefacts are not visible in the absorption contrast images in Fig. \ref{fig:fadpanel}, since there they are below the level of the noise. Reference labels delineated in coronal sections are observed at the level of the frontal cortex (FCx) in images \ref{fig:siprpanel}(a) and \ref{fig:siprpanel}(g), and level of the parietal cortex (PCx) in \ref{fig:siprpanel}(d), showing the caudate nucleus (CN), hippocampus (Hip), thalamus (Th) and corpus callosum (CC).}
  \label{fig:siprpanel}
\end{figure}

The full volume of the \textit{in situ} rabbit kitten data set was reconstructed using filtered back projection (FBP) and rotated to create axial, sagittal, and coronal views for both absorption contrast and phase retrieved PBI CT  ($\Delta\delta/\Delta\mu = 5.66 \times 10^{-10}$). Several slices of the absorption contrast CT volume can be seen in Fig. \ref{fig:fadpanel}. The views are denoted in blue, red, and green for axial, sagittal, and coronal, respectively, and the crosshairs on each panel correspond to the locations from which the other two views in that row are cut. Note that all of the images in Fig. \ref{fig:fadpanel} are conspicuously featureless apart from some bone. However, in the corresponding phase retrieved views in Fig. \ref{fig:siprpanel}, grey and white matter boundaries are resolved, and several specific brain features are clearly delineated. The overall SNR gain with phase contrast brain CT was found to be 19.7$\pm$1.5. This was determined using six of the flattest regions in each image, where the SNR is the ratio of the mean to the standard deviation of each region. This number may seem surprisingly small, given the marked improvement between Figs. \ref{fig:fadpanel} and \ref{fig:siprpanel}; however, since the radiation dose required is inversely proportional to the square of the gain\cite{Kitchen2017}, we can see that $\sim$400 times more dose would be required to obtain the same result with conventional absorption contrast CT.

Remarkably, the brain structures visible in these images are not obscured by streak or ring artefacts. This is due to the volume being rotated with respect to the CT acquisition plane in which the artefacts are created; hence the artefacts are minimised. In panel \ref{fig:siprpanel}a, the frontal lobe, frontal cortex, and striatum can be seen. Panel \ref{fig:siprpanel}d shows the parietal cortex, hippocampus and thalamus, and in panel \ref{fig:siprpanel}g the frontal cortex, corpus callosum, and caudate nucleus are all clearly resolved. The strongest streak artefacts can be seen in the axial images. This is because the axial orientation corresponds to only slightly acute angles with respect to the acquisition plane, while the angles of the other orientations are much larger.

In general, streak artefacts pose a particular problem for brain PCXI-CT due to the very low contrast between structures within the brain. Streaks that are not distinguishable above the noise in absorption contrast CT can dominate brain structures once that noise has been suppressed on phase retrieval, since they often display similar or even higher contrast. The many possible causes of these artefacts can also be very difficult to decouple. Nevertheless, we find that by paying particular care to the orientation of the sample in the CT acquisition plane, we can minimise these effects. Some residual streak artefacts still persist, and further investigation is required to determine the most appropriate method by which to correct for these effects. For preclinical studies (e.g. tissues from deceased animals or brief terminal experiments under anaesthesia) where dose is not an issue, or for studies involving non-biological samples, it should be possible to eliminate all of these artefacts by acquiring two or three CTs in orthogonal orientations.

Ring artefacts can also be seen in the lower part of the coronal images of Fig. \ref{fig:siprpanel}. As with the streak artefacts, these are minimised by sectioning slices at an angle with respect to the CT acquisition plane. For this volume, the coronal orientation is offset by 40$^\circ$ from the acquisition plane. This means that fewer consecutive image pixels contain ring artefacts originating from the same detector elements, reducing structure in the artefacts. The artefacts, however, persist to some degree along the readout direction of the detector. This results in a diffuse band that can be seen across panel \ref{fig:siprpanel}b from the lower centre of the image, running diagonally upward and to the right, through the cerebellum. This effect is most prominent at the centre of rotation of the sample and becomes less so at larger radii.

It should also be noted that it is of particular importance for phase contrast brain imaging to accurately account for all of the phenomena that cause each of the different types of artefact. There are many correction methods that work effectively for absorption contrast CT imaging that are insufficient for the low-contrast edges that are enhanced using phase contrast imaging, as the latter exposes effects of these artefacts that, while present in absorption contrast imaging, are not generally problematic as they are below the noise threshold. With further development of artefact correction methods (e.g. PSF deconvolution, improved detector characterisation, modelling source variations, etc.), we anticipate that the SNR and CNR of phase contrast brain CT can be even further improved.

\section*{Conclusions}

Propagation-based PCXI-CT is shown here to be an effective tool for visualising the brain \textit{in situ} for preclinical animal studies. While the surrounding skull, temporal fluctuations in the intensity of the source, and detector imperfections provide distinct challenges with respect to reconstruction artefacts, we find that there are ways to work around these limitations to see brain structures that might otherwise be obscured. We present a two-material phase retrieval algorithm for tomography, which was shown to be highly effective at delineating soft tissue from bone. In addition, we find that by changing the `sectioning angle' of the 3D volume, we are able to significantly reduce contamination from streak and ring artefacts. We have identified the most problematic causes of these artefacts, and while further work will be required to address these phenomena, it is clear that it is already possible to identify structures that have previously been unresolved with conventional X-ray CT. The substantial SNR gain achieved using PCXI-CT has no requirement for contrast agents and allows for the visualisation of features that would otherwise require a $\sim$400-fold increase in the radiation dose required to obtain the equivalent results with conventional absorption contrast CT.

\section*{Methods}
\label{sec:experiments}

This experiment used rabbit kittens that had been used in experiments conducted with approval from the SPring-8 Animal Care (Japan) and Monash University (Australia) Animal Ethics Committees. All experiments were performed in accordance with relevant guidelines and regulations. The kittens were humanely killed in line with approved guidelines and the carcasses scavenged for this experiment.

To examine CT streak artefacts from strongly-absorbing samples, simulations were performed of a CT of an aluminium rod with a length and thickness designed to mimic those of the rabbit kitten skulls in the CTs discussed below. For comparison, a CT was experimentally acquired of an aluminium phantom at 26 keV at the shortest feasible propagation distance of 12 cm. The phantom consisted of a 0.3 mm thick strip of 99\% pure aluminium sheeting suspended in agarose and was specifically designed to aid determination of the primary cause of the streak artefacts seen in \textit{in situ} brain imaging by mimicking the projected thickness of the highly attenuating skull. The phantom CT was acquired using a 4000 $\times$ 2672 pixel Hamamatsu CCD camera (C9300-124) with a tapered fiber optic bonded between the CCD chip and the \SI{20}{\um} thick gadolinium oxysulfide (Gd$_2$O$_2$S; P43) phosphor, with an effective pixel size of \SI{16.2}{\um}. Each CT consisted of 1800 projections spanning 180$^{\circ}$ of rotation, with an exposure time of 80 ms per projection.

To visualise the brain, CTs were acquired of a scavenged New Zealand White rabbit kitten head and an excised rabbit kitten brain, both at 30 days gestational age (GA; term $\sim$32 days), suspended in agarose. They were acquired at an energy of 24 keV and at a 5 m sample-to-detector propagation distance using a 2048 $\times$ 2048 Hamamatsu digital sCMOS camera (C11440-22C) with a \SI{25}{\um} thick gadolinium oxysulfide scintillator and a pixel size of \SI{15.1}{\um}. Further CTs were acquired at a higher resolution in order to test the effects of detector resolution on streak artefacts. These consisted of a rabbit kitten brain \textit{in situ}, excised from a scavenged New Zealand White rabbit kitten at 29 days GA, suspended in agarose, at propagation distances of 5 m and 11 cm. Both were captured using a second 2048 $\times$ 2048 Hamamatsu digital sCMOS camera (C11440-22C) with a straight fibre optic and a \SI{15}{\um} thick gadolinium oxysulfide scintillator with a pixel size of \SI{6.49}{\um}. Due to the projected sample size being larger than the detector field of view, 7200 projections were taken through 360$^{\circ}$ of rotation, which were later stitched together with linear blending to create a single dataset of 3600 projections spanning 180$^{\circ}$, with a dose rate of 33.9 mGy/s.

All CTs were acquired at a source-to-object distance of 210 m on beamline BL20B2 at the SPring-8 synchrotron in Japan and were reconstructed using FBP. Reconstructions were performed on the MASSIVE supercomputer in Melbourne, Australia using the ASTRA Toolbox CUDA accelerated FBP algorithm\cite{VanAarle2015}.

\bibliography{references}

\section*{Acknowledgements}

We thank Erin McGillick for her assistance in procuring supplies necessary for these experiments. L.C.P.C. is supported by an RTP scholarship. M.J.K. is supported by an ARC Future Fellowship (FT160100454). S.B.H. is an NHMRC Principal Research Fellow. K.S.M was supported by a Veski VPRF and completed this work with the support of the TUM Institute for Advanced  Study,  funded  by  the  German  Excellence  Initiative  and  the  European  Union  Seventh Framework Program under grant agreement no 291763 and co-funded by the European Union. This work was funded by ARC Discovery Project DP170103678 and supported by SPring-8 proposals 2015B0047 and 2016A0047. We acknowledge travel funding provided by the International Synchrotron Access Program (ISAP) managed by the Australian Synchrotron and funded by the Australian Government (AS/ IA153/10571). This work is supported by the Victorian Government's Operational Infrastructure Support Program. Data analyzed in this paper have been deposited into the Store.Monash repository, identifiable by the doi:10.4225/03/5a56e4df15309.

\section*{Author contributions statement}

L.C.P.C., M.J.K., M.J.W., L.T.K., and S.B.H. conducted the experiment with assistance and expertise from K.U. and N.Y.; analysis of data by L.C.C.P and M.J.K.; theoretical contributions by L.C.C.P, M.J.K, K.S.M., and D.M.P.; physiological contributions by S.B.H., M.J.W., L.T.K., K.J.C. and S.L.M.; the manuscript was written by L.C.C.P, M.J.K., K.S.M., and D.M.P. and edited by all authors.

\end{document}